\documentclass{PoS}

\newcommand{\as}{\alpha_{\rm s}}

\def\nl{{n^{}_{\! l}}}
\def\nh{{n^{}_{\! h}}}

\title{Mass effects and four-particle amplitudes at the two-loop level
 in QCD}

\ShortTitle{Mass effects and four-particle amplitudes at the two-loop
 level in QCD}

\author{\speaker{Michal Czakon}%
         \thanks{This work was supported by the Sofja Kovalevskaja Award of the
           Alexander von Humboldt Foundation sponsored by the German Federal
           Ministry of Education and Research.}\\
        Institut f\"ur Theoretische Physik
        und Astrophysik, Universit\"at W\"urzburg, \\
        Am Hubland, D-97074 W\"urzburg, Germany \\
        Institute of Physics, University of Silesia, \\
        Uniwersytecka 4, PL-40007 Katowice, Poland\\
        E-mail: \email{mczakon@physik.uni-wuerzburg.de}}

\abstract{I review the recent progress in the evaluation of two-loop
  massive QCD amplitudes necessary for the description of hadronic scattering
  processes at the LHC. I put a special emphasis on the case of top-quark
  final states.}

\FullConference{8th International Symposium on Radiative Corrections \\
		 October 1-5, 2007\\
		 Florence, Italy}

\begin{document}

\section{Introduction}

There are just a few processes, yet to be observed at the LHC, which require a
true NNLO analysis matching a foreseen experimental precision at the percent
level. One of the most notable such processes is top-quark pair
production. Since the current NLO prediction has a scale dependence of the
order of 13\% \cite{Nason:1988xz} and soft-gluon resummations
\cite{Laenen:1991af,Bonciani:1998vc,Kidonakis:2003qe} are not reliable, the
case for conducting a complete analysis at the next order of perturbation
theory is well grounded.

Here, I review results, which have been obtained thanks to recent advances in
the understanding of the application of the factorization theorem to collinear
limits \cite{Mitov:2006xs} on the one hand, and to progress in the evaluation
of massive amplitudes started in \cite{Czakon:2004wm,Czakon:2006pa} on the
other. By combining the methods, the leading asymptotics in the high-energy
regime of the virtual amplitudes have been derived in
\cite{Czakon:2007ej,Czakon:2007wk}. From then on, the direct evaluation
approach has also provided power corrections in the mass and fully numerical
solutions that cover the whole kinematically interesting range
\cite{CzakonNew1}.

\section{Leading asymptotics}

Based on experience, it is relatively clear that the logarithmic terms of the
expansion of any observable in any limit should be much easier to obtain than
the full result. How to actually derive them in the case of collinearly
singular limits has been first worked out for Bhabha scattering
\cite{Glover:2001ev}. Later on, it turned out that even the constant should be
within reach thanks to the use of factorization properties of amplitudes
\cite{Penin:2005kf}. For QCD, the full procedure has been presented in
\cite{Mitov:2006xs} (for application in the context of Bhabha scattering see
\cite{Becher:2007cu}). Until now, the only part that poses problems is the one
due to contributions of closed heavy particle loops, since they actually
generate soft singularities, which are process dependent in the sense, that
they involve a non-trivial color flow between different external legs.

Let us expand the top-quark pair production amplitude in the strong
coupling constant as follows

\begin{eqnarray}
  \label{exp}
  | {\cal M} \rangle
  & = &
  4 \pi \as \biggl[
  | {\cal M}^{(0)} \rangle
  + \biggl( {\as \over 2 \pi} \biggr) | {\cal M}^{(1)} \rangle
  + \biggl( {\as \over 2 \pi} \biggr)^2 | {\cal M}^{(2)} \rangle
  + {\cal O}(\as^3)
  \biggr]
\, .
\end{eqnarray}

If the production process involves quark pair annihilation, one can perform
the following color decomposition of color averaged amplitudes

\begin{eqnarray}
  \label{quark}
{\lefteqn{
  2 {\rm Re}\, \langle {\cal M}^{(0)} | {\cal M}^{(2)} \rangle = }}
\\
& &
2 (N^2-1) \biggl(
N^2 A + B  + {1 \over N^2} C
+ N \nl D_l + N \nh D_h
+ {\nl \over N} E_l + {\nh \over N} E_h + (\nl+ \nh)^2 F
\biggr)
\, ,
\nonumber
\end{eqnarray}

where $N$ is the number of colors, $\nh$ the number of heavy quarks and $\nl$
the number of light quarks.  As noted above, all terms which are free of
heavy-quark loops, i.e. $A, B, C, D_l, E_l$ can be obtained from
factorization. Although the term $F$ involves heavy quark loops, it is simple
enough to be inferred analogously to the remaining ones. In principle, to
complete the expression, one would only have to compute $D_h$ and $E_h$. These
contributions are particularly easy to evaluate with the help of methods based
on Mellin-Barnes representations
\cite{Smirnov:1999gc,Tausk:1999vh,Anastasiou:2005cb,Czakon:2005rk}. The results for all the
coefficients have been presented in \cite{Czakon:2007ej}, where the authors
also checked the factorization formalism by computing directly the $A$, $D_l$
and $F$ coefficients finding, as expected, perfect agreement. The interested
reader is encouraged to consult \cite{Czakon:2007ej} for a detailed
description of the techniques used. Here, I should only stress that the
evaluation of the subleading color coefficients is substantially more
complicated due to the occurrence of non-planar graphs. In the mean time,
however, all the coefficients have been obtained by the present author,
confirming the published results.

Similarly to the quark annihilation channel, the averaged amplitude in the
gluon fusion channel can be decomposed as follows

\begin{eqnarray}
  \label{gluon}
{\lefteqn{
  2 {\rm Re}\, \langle {\cal M}^{(0)} | {\cal M}^{(2)} \rangle =
(N^2-1) \biggl(
N^3 A + N B  + {1 \over N} C + {1 \over N^3} D
+ N^2 \nl E_l + N^2 \nh E_h
}}
\\
& &
+ \nl F_l + \nh F_h
+ {\nl \over N^2} G_l + {\nh \over N^2} G_h
+ N \nl^2 H_l + N \nl \nh\, H_{lh} + N \nh^2 H_h
+ {\nl^2 \over N} I_l + {\nl \nh \over N} I_{lh} + {\nh^2 \over N} I_h
\biggr)
\, .
\nonumber
\end{eqnarray}

The results for all the coefficients have been given in
\cite{Czakon:2007wk}. The direct  computation covered $A, E_l, E_h, F_h, G_h,
H_l, H_{lh}, H_h, I_l, I_{lh}, I_h$, whereas the factorization formalism
provided $A$, $B$, $C$, $D$, $E_l$, $F_l$, $G_l$, $H_l$, $H_{lh}$, $H_h$,
$I_l$, $I_{lh}$, $I_h$. Note that this time it  was impossible not to compute
non-planar graphs, since part of the heavy quark loop contributions involved
them. The direct computation of the coefficients covered by factorization only
is still under way, since it is necessary for the evaluation of the full
amplitude.

\section{Power corrections}

Due to parton fluxes, a substantial part of top quark pairs will be
produced relatively close to threshold. It should, therefore, be clear that the
high-energy asymptotics described in the previous section is insufficient for
practical purposes. An exact evaluation of the amplitude seems to be out of
reach at present, but one can expand in the mass. Introducing
\begin{equation}
  s = (p_1+p_2)^2, ~~~~ t = (p_1-p_3)^2-m_t^2, ~~~~ x = -\frac{t}{s},
\end{equation}
where $p_1$ and $p_2$ are the momenta of the incoming partons and $p_3$ is the
momentum of the outgoing top-quark, and
\begin{equation}
  x \in [ 1/2(1-\beta), 1/2(1+\beta) ], ~~~~ \beta = \sqrt{1-4m_t^2/s}.
\end{equation}

With $r=4m_t^2/s$ and $x=1/2$, which corresponds to 90 degree scattering,
the expansion of the bare leading color term $A$ is
\begin{equation}
\includegraphics[width=14cm]{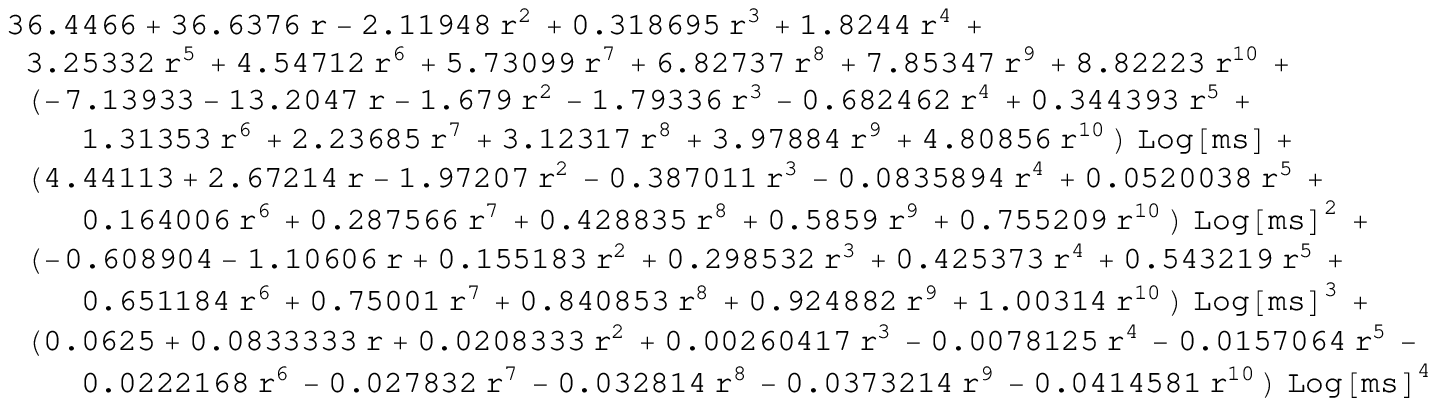}
\end{equation}

\begin{figure}
\begin{center}
\includegraphics[width=7cm]{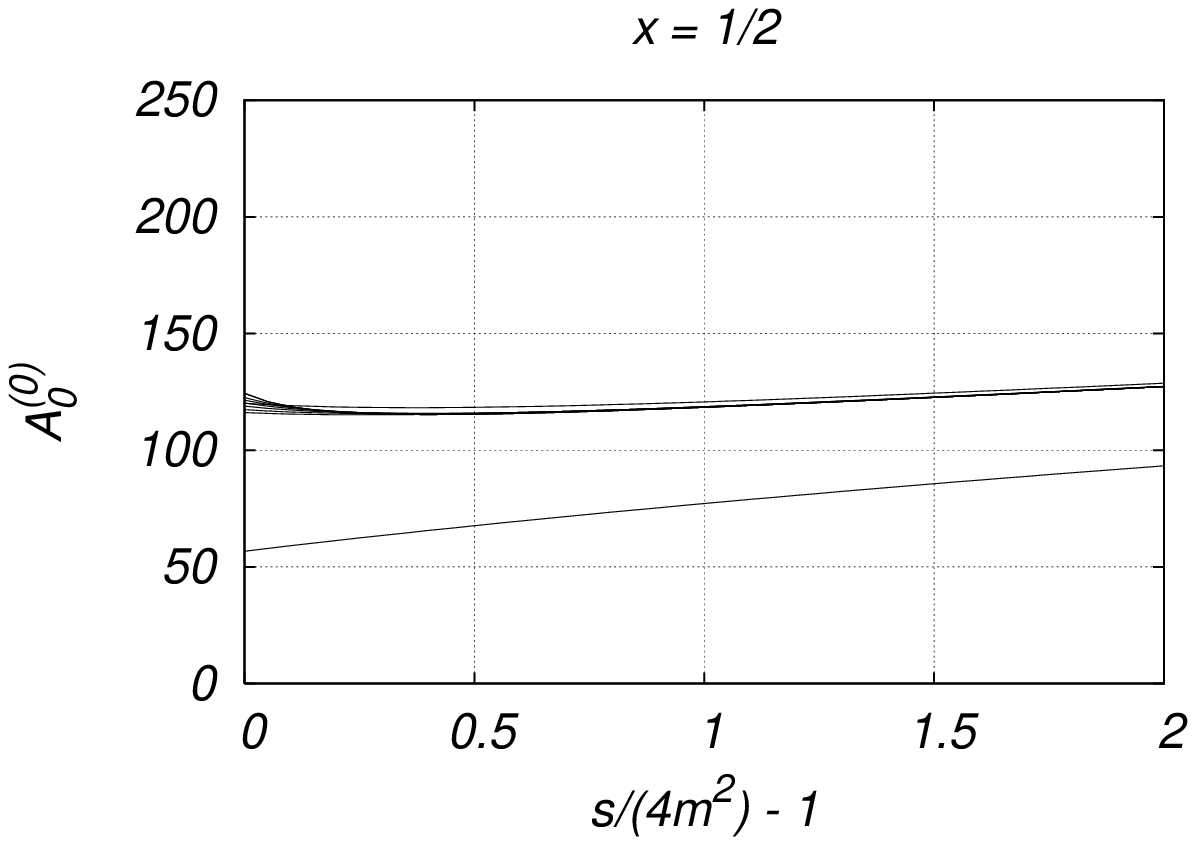}
\includegraphics[width=7cm]{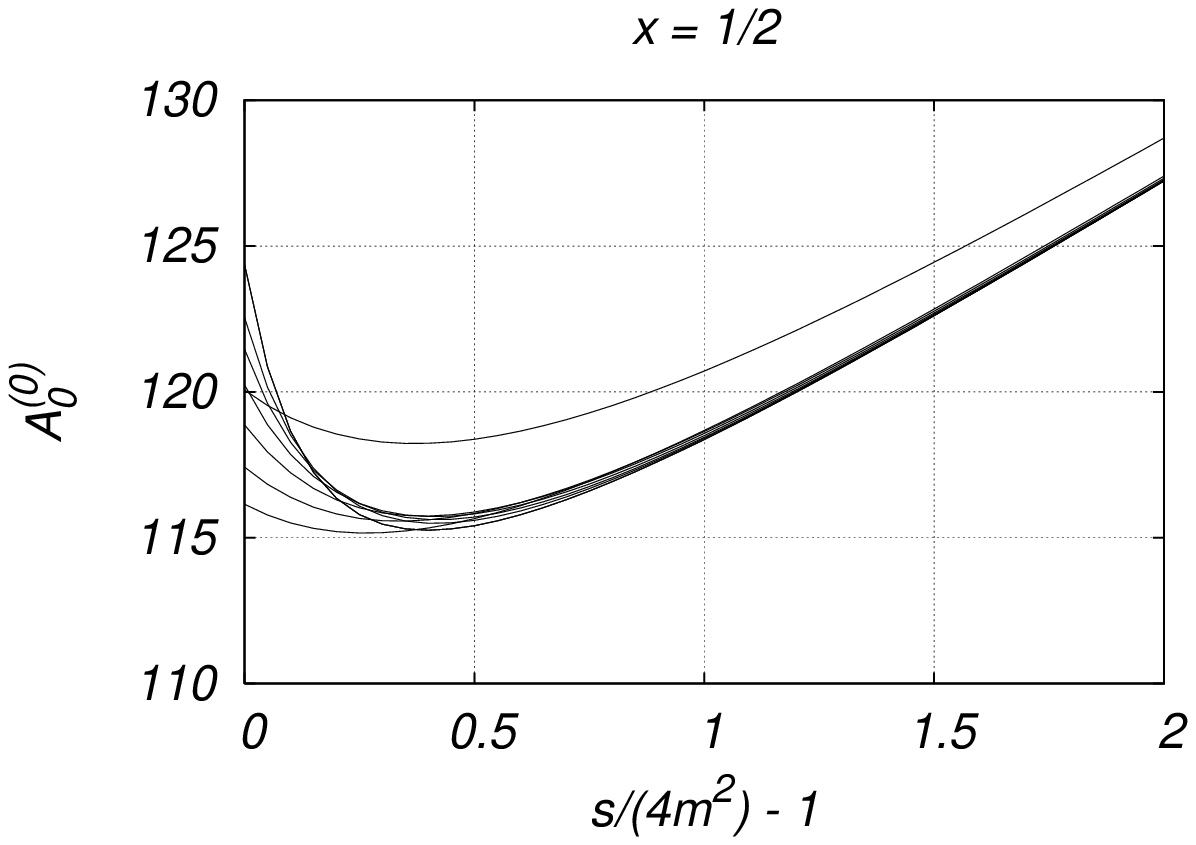}
\includegraphics[width=7cm]{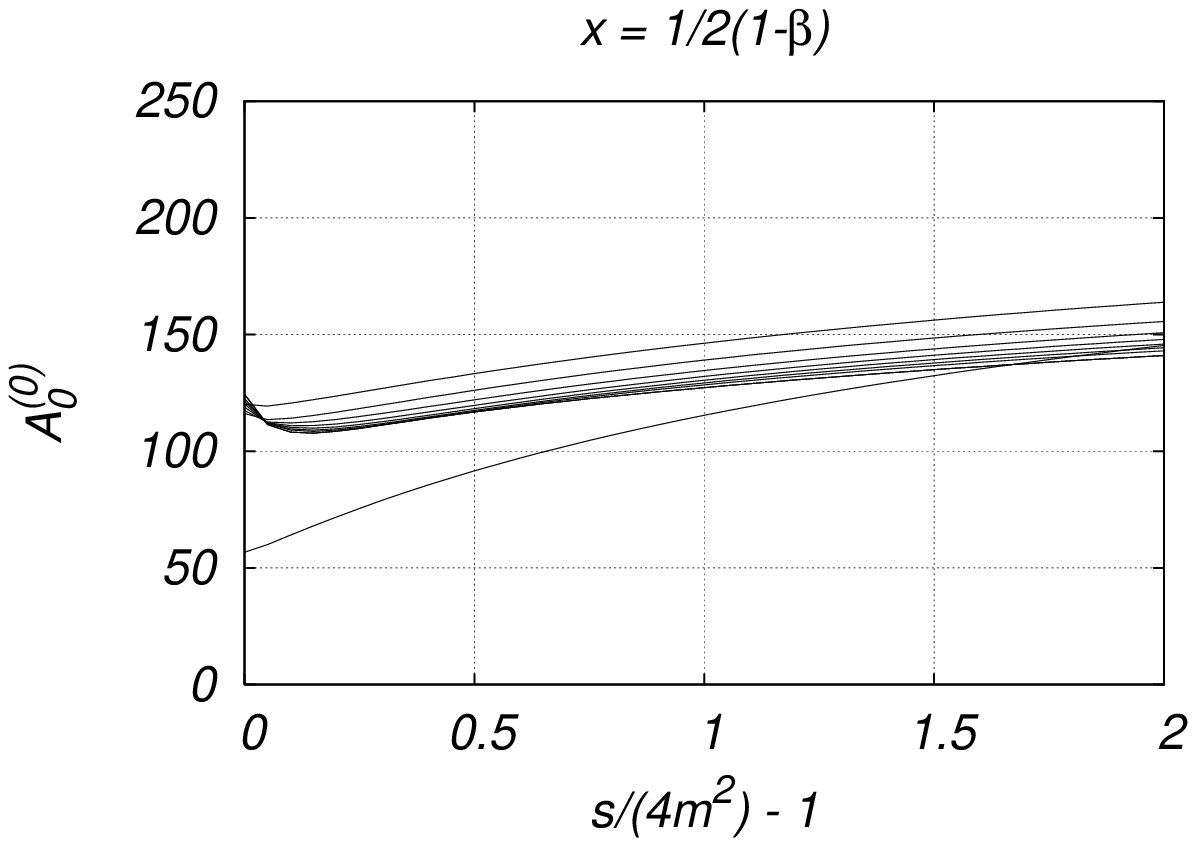}
\includegraphics[width=7cm]{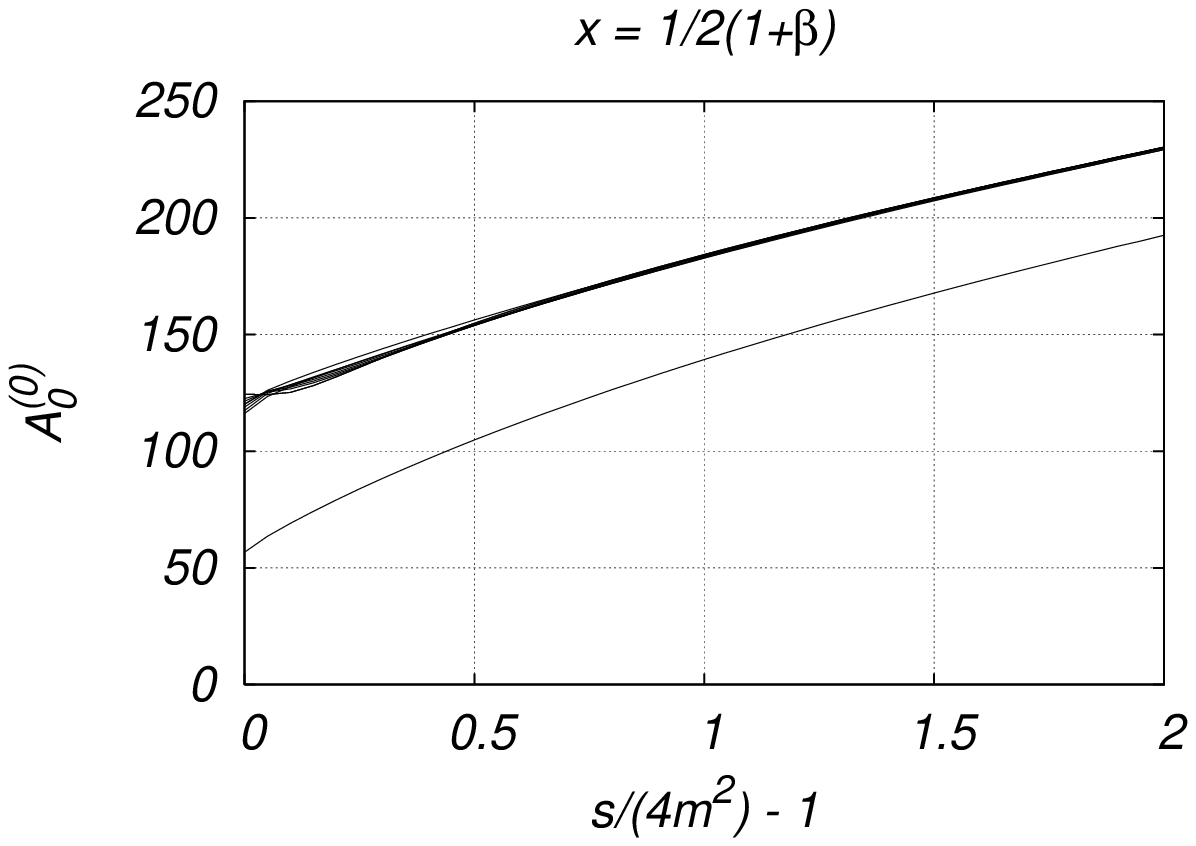}
\caption{Bare leading color two-loop amplitude for top quark pair production
  in the quark annihilation channel expanded in the mass. The more divergent
  terms at threshold (left of the plots) correspond to higher orders of
  expansion. The variables are defined in the text.}
\label{czakon_expansion}
\end{center}
\end{figure}

This result in itself looks promising, because we observe asymptotic
convergence of the series even at threshold. Unfortunately, the subleading
color coefficients cannot behave so nicely, as they contain Coulomb
singularities. Moreover, the series suffers from convergence problems for
small and large angles as depicted in Fig.~\ref{czakon_expansion}.

Further discussion and details can be found in \cite{CzakonNew1}.

\section{Numerical solutions}

\begin{figure}
\begin{center}
\includegraphics[width=12cm]{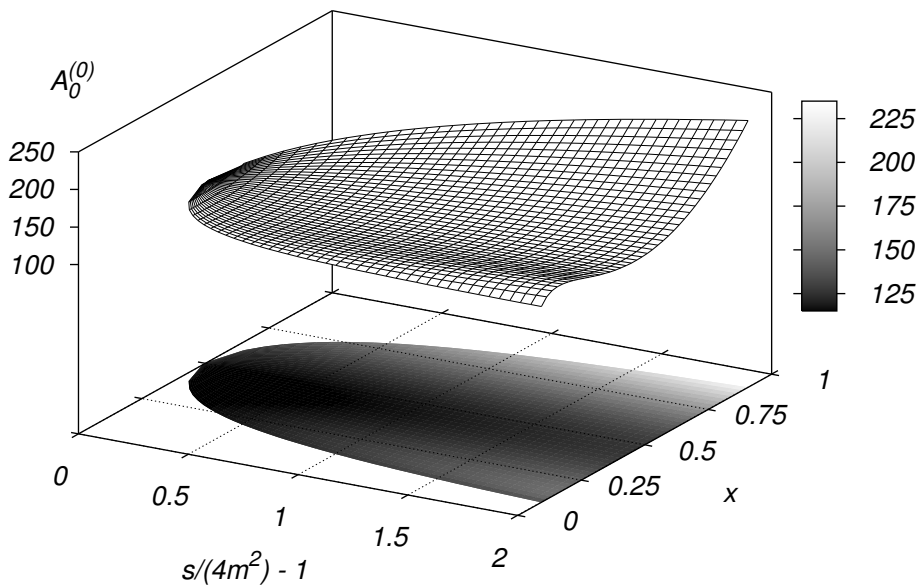}
\caption{Full mass dependence of the bare leading color amplitude in the quark
annihilation channel.}
\label{czakon_full}
\end{center}
\end{figure}

As the discussion of the previous section shows, it is impossible to proceed
further without evaluating the amplitude in the threshold region. It turns out
that this can be done efficiently with the method of differential equations
\cite{Kotikov:1990kg,Remiddi:1997ny,Caffo:2002ch} (for previous applications
to physical processes see \cite{Boughezal:2007ny,Czakon:2007qi}). The results
for the bare leading color coefficient are shown in
Fig.~\ref{czakon_full}. For the details the reader is deferred again  to
\cite{CzakonNew1}.

\section{Conclusions}

Although there is still some work to be done before obtaining a true NNLO
prediction for the total cross section for the top-quark pair production cross
section, we are at the stage where the virtual amplitudes are almost known
(notice that the one-loop squared contributions are known since
\cite{Korner:2005rg,Korner:2008bn}). The real-virtual corrections are mostly
known thanks to the  $t\bar{t}$+jet calculation \cite{Dittmaier:2007wz} and
the necessary phase space integrations of the real radiation  corrections can
be done with the help of sector decomposition
\cite{Binoth:2000ps,Binoth:2004jv}, which would then complete the program. Let
me also note that similar progress has been obtained in the case of
gauge-boson pair production \cite{CzakonNew2} (here, the gauge boson pair
production with an additional jet has been studied very recently in
\cite{Dittmaier:2007th,Campbell:2007ev}).

\end{document}